\documentclass[prb,twocolumn,showpacs,floatfix,amsmath]{revtex4-1}

\usepackage{color}
\usepackage{xcolor}
\usepackage{dsfont}
\usepackage[normalem]{ulem}
\usepackage{dcolumn}

\usepackage{graphicx}
\def\be{\begin{equation}}
\def\ee{\end{equation}}
\def\beq{\begin{eqnarray}}
\def\eeq{\end{eqnarray}}

\begin{document}
\renewcommand{\familydefault}{\sfdefault}
\renewcommand{\sfdefault}{cmbr}
\title{Impact of single atomic defects and vacancies on the magnetic anisotropy energy of CoPt thin films}
\author{Samy Brahimi$^1$}\email{samy86brahimi@gmail.com}
\author{Hamid Bouzar$^1$}
\author{Samir Lounis$^2$}
\affiliation{$^1$ Laboratoire de Physique et Chimie Quantique, Universit\'e Mouloud Mammeri de Tizi-Ouzou, 15000 Tizi-Ouzou, Algeria}
\affiliation{$^2$ Peter Gr\"unberg Institut and Institute for Advance Simulation, Forschungszentrum J\"ulich \& JARA, 52425 J\"ulich, Germany}

\begin{abstract}
The impact of surface vacancies and single adatoms on the magnetic properties of  tetragonal {\bf{L1}$_{0}$} CoPt thin films is investigated from first principles. We consider Co and Fe single adatoms deposited on a Pt-terminated thin film while a Pt adatom is assumed to be supported by a Co-terminated film. The vacancy is injected in the top-surface layer of the films with both types of termination. After finding the most stable location of the defects, we discuss their magnetic properties tight to those of the substrate and investigate the magnetic crystalline anisotropy energy (MAE). Previous simulations [Brahimi et al. J. Phys.: Condens. Matter. \textbf{28}, 496002 (2016)] predicted a large out-of-plane surface MAE for the Pt-terminated CoPt films (4 meV per f.u.) in contrast to in-plane surface MAE for  Co-terminated films (-1 meV per f.u.). 
Here, we find that the surface MAE is significantly modified upon the presence of the atomic defects. All investigated defects induce an in-plane MAE, which  is large enough for Fe adatom and Pt vacancy to switch the surface MAE from out-of-plane to in-plane for the Pt-terminated films. Interestingly, among the investigated defects Pt vacancy has the largest effect on the MAE in contrast to Co vacancy, which induced the smallest but still significant effect. This behavior is explained in terms of the orbital moment anisotropy of the thin films.
\end{abstract}      

\maketitle
\date{\today}

\section{Introduction}
The magnetoscrystalline anisotropy energy (MAE) has a prominent role in the realm of information technology that depends on storing data in terms of magnetic bits. Indeed, the MAE defines the energy barrier that stabilizes the direction of magnetic moments.  In this context, the search for materials exhibiting large MAE, in particular those with a perpendicular magnetization, have attracted a lot of interest because of their several possible applications in magneto-optical storage, magneto-electronic devices and spintronics in general. ~\cite{mccallum2014practical, Iwasaki1980,Victora2005,bader2002magnetism,Gu2014,Chappert2007,Dieny2017}.

Bulk CoPt binary alloy in the {\bf{L1}$_{0}$} structure  is known for its large uni-axial MAE, of around 1 meV~\cite{Eurin1969,Maykov1989,Ye1990,Weller1999,Eurin1969,Grange2000,Galanakis2000,Ravindran2001,Schick2003,Yang2004,Perez2005}, which is favored by its tetragonal distortion (see e.g. Refs.~\cite{Sakuma1994,Razee2009,brahimi2016giant}). Interestingly, when considered in the film-geometry along the [001] direction, the same alloy was predicted  to have a large surface contribution to the MAE  when the film is terminated with a Pt surface layer~\cite{brahimi2016giant} in contrast to the Co-terminated films~\cite{Zhang2009,brahimi2016giant}. Besides the (001) surface, we note that several other surfaces of CoPt thin films were recently investigated theoretically~\cite{Liu2017}.

The goal of our current study is to investigate the stability of the MAE of CoPt films against the presence of atomic defects or vacancies, which constitutes an extension of our previous study, where we considered  extended defects, such as stacking faults and anti-sites defects.  Taken from another perspective, our study is interesting in the context of magnetism of nanostructures down to single atoms being at the vicinity of surfaces, wherein one of the prospected aims is to explore the possibility of stabilizing and manipulating  single atomic magnetic bits. We note that since the seminal work of Gambardella et al.~\cite{gambardella2003giant} demonstrating large perpendicular MAE characterizing Co adatoms deposited on Pt(111) surface, reaching 10 meV, several theoretical and experimental studies have been performed to explore the MAE, among various properties, of single atomic defects on different non-magnetic metallic substrates and in particular on Pt (see e.g.~\cite{Etz2008,Balashov2009,Khajetoorians2013,Dubout2015,Bouhassoune2016}). 
The choice for Pt(111) surface was motivated by its strong spin-orbit interaction carried by the 5$d$ element Pt, which can enhance the magnitude of the MAE. The CoPt surface is in this context interesting since it can also be terminated by a Pt layer which is already magnetic because of the underlying Co layers. To our knowledge, no investigation related to the MAE has been devoted to the case of adatoms deposited on such alloys. Furthermore, while the investigations on non-magnetic substrates focus on the MAE of the adatoms, we aim at scrutinizing how the low doping of magnetic impurities can modify the MAE of the whole substrate. 

We consider single atomic defects such as Co and Fe adatoms  supported by Pt-terminated (001) thin CoPt films (see Fig.\ref{cell_surf}(c-e-d)) and Pt adatom on the  Co-terminated ones (see Fig.\ref{cell_surf}(c-e-d)).  Our ab-initio based simulations indicate that among the three types of adatoms, Fe is found to induce the largest in-plane MAE. Besides adatoms, we study single Co or Pt  vacancies (see Fig.\ref{cell_surf}(a-b)).  Similarly to the adatoms, we find that vacancies favor an in-plane MAE of the films reaching a maximum when the Pt vacancy in injected on the surface.  The article is organized as follows. First we provide a description of the utilized method and numerical considerations, which is followed by the results section. In the latter we address the structurall relaxations induced by the defects and determine the most preferable location of the adatoms. The impact of the adatoms and vacancies on the magnetic properties of the substrate is studied, which is followed by the investigation and analysis of the MAE.  Finally, a conclusion summarizing our results is provided.

\section{Method}
We simulate the CoPt thin films by adopting the slab approach with periodic boundary 
conditions in two directions while the periodic images in the third direction are 
separated by a sufficient amount of vacuum (15 \AA) to avoid  interaction between 
neighboring supercells. Unit cells with an odd number of planes  (9 layers) are chosen to avoid the pulay stress. The number of atoms per layer in each unit cell is four for the different studied cases as can be seen in some representative slabs shown in Fig.\ref{cell_surf}. 
Here it can be observed that for equiatomic {\bf{L1}$_{0}$} type of alloys two different 
surfaces, made of either Co or Pt, exist when the slabs are stacked along the {[001]} direction. On the one hand, single Co and Fe adatoms are deposited on the Pt-terminated substrate while Pt adatom is considered on top of Co-terminated surface. On the other hand vacancies are injected in the surface layer of both substrates. 
The self-consistent calculations are carried out with the Vienna ab initio simulation package (VASP) 
using a plane wave basis and the projector augmented wave (PAW) approach ~\cite{PAW1,PAW2}. 
The exchange-correlation potential is used in the functional from of Perdew, Burke and Ernzerhof (PBE) ~\cite{PBE1,PBE2}. 
The cut-off energies for the plane waves is 478 eV. The integration over the Brillouin zone was based on finite temperature smearing (Methfessel-Paxton method) with a  k-points grid of \(14\times14\times1\) points. The energy 
convergence criterion is set to \(10^{-8}\) eV while the geometrical atomic relaxations for the surfaces calculations were stopped when the forces were less than 0.01 eV/\AA. 
The MAE is extracted from the difference between the total energies of the two configurations: out-of-plane versus in-plane orientations of the 
magnetic moments. A positive value indicates a preference for the 
out-of-plane orientation of the magnetic moments.

\section{Results}
\subsection{Structural relaxations and magnetic moments}
\begin{figure*}[htpb]
\centering
\includegraphics[width=2\columnwidth]{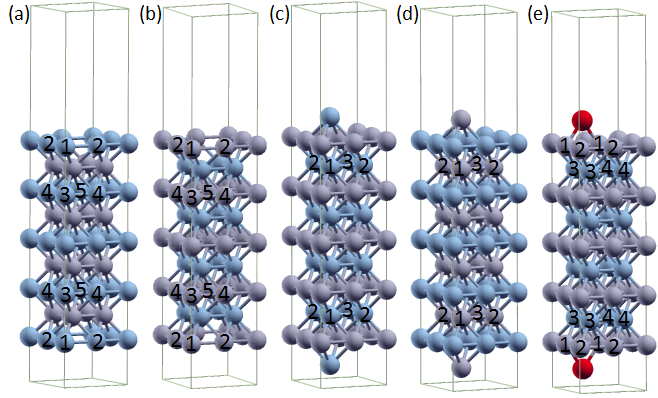}
\caption{Supercells used for the simulation of the (001) CoPt surface with different types of single atomic defects where the blue, 
magenta and red spheres correspond respectively 
to the Co, Pt and Fe atoms: (a) Surface with Co vacancy, (b) Surface with Pt vacancy, (c) Co adatom, (d) Pt adatom and (e) Fe adatom. The numbers 1, 2, 3, 4 and 5 refer to atoms with different magnetic moments.}
\label{cell_surf}
\end{figure*}

\textbf{Vacancies.}
As mentioned previously, we inject vacancies, with a concentration of 25\%, in the top-surface layer of the CoPt alloy by removing either a Co or a Pt atom depending on the surface termination (see Fig.~\ref{cell_surf}(a-b)). We provide in Table~\ref{table_perfect} the atomic magnetic moments obtained after full atomic relaxations. The atoms labeled 1 and 2 relax differently along the z-direction leading to a sort of roughness, since the vacancy breaks the intra-layer symmetry. Introducing the vacancy provides enough space for the surface atoms to relax inward. Interestingly, the relaxation is much larger when creating the vacancy in the Pt surface layer than in the Co surface layer.  In the case of the Pt vacancy, atoms 1 and 2 relax respectively by -13.64 \% and -10.86 \% comparatively to the bulk  interlayer distance while for the Co vacancy  relaxations are more moderate, -7.54 \% and -8.50 \% for respectively atom 1 and 2. We recall that for the perfect surface~\cite{brahimi2016giant}, the surface layers relaxed by only -5,60\% and -4.30\% for respectively the Pt- and Co-terminated surfaces.
The subsurface layer is also affected by the missing surface atom, which introduces geometrical rearrangements of the atoms being more moderate for the case of a Co vacancy (-1.12 \%) than for the case of a Pt vacancy (-7.06 \%). 

By scrutinizing the magnitude of the spin moments, we found that surface Pt atoms surrounding the vacancy experience a slight enhancement of the moments. For instance Pt atoms, labeled as atoms 2, display a moment of 0.49 $\mu_{B}$ as indicated in Table~\ref{table_perfect} instead of 0.42 $\mu_{B}$ obtained on the perfect surface. Although the layer below the subsurface layer did not relax geometrically, the magnetic moments can be strongly modified by the presence of surface vacancies.

\begin {table*}[htpb]
\begin{center}\begin{tabular}{|c|c|c|c|}
		\hline
		\multicolumn{3}{|c|}{Co-vacancy}\\
		\hline
\hline
	Layer & Atom & M ($\mu_{B}$)\\\hline
	S & ~\begin{tabular}{c}
	Co$_{1}$\\
	Co$_{2}$\\	
\end{tabular}
 & ~\begin{tabular}{c}
	1.99\\
	2.01\\	
\end{tabular}
\\\hline
	S-1 & ~\begin{tabular}{c}
	Pt\\
\end{tabular}
 & ~\begin{tabular}{c}
	0.41\\
\end{tabular}
\\\hline
	S-2 & ~\begin{tabular}{c}
	Co$_{3}$\\
	Co$_{4}$\\
	Co$_{5}$\\
\end{tabular}
 & ~\begin{tabular}{c}
	1.98\\
	1.88\\
	1.85\\
\end{tabular}
\\\hline
	S-3 & ~\begin{tabular}{c}
	Pt\\
\end{tabular}
 & ~\begin{tabular}{c}
	0.40\\
\end{tabular}
\\\hline
	Center & ~\begin{tabular}{c}
	Co\\
\end{tabular}
 & ~\begin{tabular}{c}
	1.93\\
\end{tabular}
\\\hline
\end{tabular}
\hspace{1cm}
\begin{tabular}{|c|c|c|c|}
			\hline
	\multicolumn{3}{|c|}{Pt-vacancy}\\
	\hline
\hline
	Layer & Atom & M ($\mu_{B}$)\\\hline 
	S & ~\begin{tabular}{c}
	Pt$_{1}$\\
	Pt$_{2}$\\
\end{tabular}
 & ~\begin{tabular}{c}
	0.40\\
	0.49\\
\end{tabular}
\\\hline
	S-1 & ~\begin{tabular}{c}
	Co\\
\end{tabular}
 & ~\begin{tabular}{c}
	2.00\\
\end{tabular}
\\\hline
    S-2 & ~\begin{tabular}{c}
	Pt$_{3}$\\
	Pt$_{4}$\\
	Pt$_{5}$\\
\end{tabular}
 & ~\begin{tabular}{c}
	0.45\\
	0.39\\
	0.41\\
\end{tabular}
\\\hline
    S-3 & ~\begin{tabular}{c}
	Co\\
\end{tabular}
 & ~\begin{tabular}{c}
	1.91\\
\end{tabular}
\\\hline
    Center & ~\begin{tabular}{c}
	Pt\\
\end{tabular}
 & ~\begin{tabular}{c}
	0.41\\
\end{tabular}
\\\hline
\end{tabular}\end{center}
\caption{Spin magnetic moments of thin films with vacancies injected in the top surface layer. }
\label{table_perfect}
\end{table*}

\begin{figure}[htpb]
\centering
\includegraphics[width=\columnwidth]{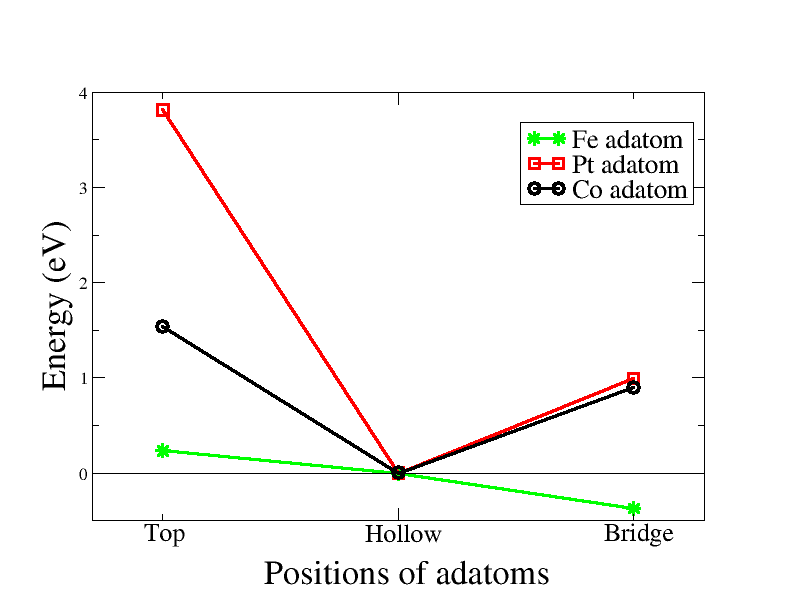} 
\caption{Relative energies of the CoPt thin films as function of the locations of Co, Pt and Fe single adatoms in top-, bridge- and hollow-configurations. The reference energy is defined by the hollow-configuration's total energy.}
\label{sites}
\end{figure}

\textbf{single adatoms.}
In the case of adatoms, we investigated the case of Co and Fe single atoms deposited on the Pt-terminated surface of CoPt and of Pt atom deposited on the Co-terminated surface (see Fig.\ref{cell_surf}(c-e)). We have explored three possible configurations that are likely to be occupied by the single atoms: top-, bridge- and hollow-sites. As can be grasped from Fig.~\ref{sites}, we found that the ground state of Co and Pt adatoms is the hollow site in contrast to the Fe adatom, which prefers the bridge-site.

 Co and Pt adatoms experience strong inward relaxations of respectively -15.88 \% and -26.42 \%. In the case of Fe adatom, which is located in the bridge-site, the surface Pt atoms relax differently, depending on their separation distance from the adatom. The "bridge" on which the Fe adatom is deposited define Pt atoms (1). On the one hand, the distance between Fe and those Pt atoms increases with respect to the bulk equivalent distance (3.48\%). On the other hand, the adatom distance to the other surface Pt atoms, denoted (2),  decreases (-8.02\%). Similarly to the case of vacancies, the subsurface layer experiences corrugation but with a much smaller magnitude. The large relaxations have necessarily an impact on the magnetic moments as shown by Table \ref{table_ad_atoms}.

\begin {table*}[htpb]
\begin{center}\begin{tabular}{|c|c|c|c|}
				\hline
		\multicolumn{3}{|c|}{Co adatom}\\
		\hline
\hline
	Layer & Atom & M ($\mu_{B}$)\\\hline
	Adatom & ~\begin{tabular}{c}
	Co\\
\end{tabular}
 & ~\begin{tabular}{c}
	2.13\\
\end{tabular}
\\\hline 
	S & ~\begin{tabular}{c}
	Pt\\
\end{tabular}
 & ~\begin{tabular}{c}
	0.41\\
\end{tabular}
\\\hline
	S-1 & ~\begin{tabular}{c}
	Co$_{1}$\\
	Co$_{2}$\\
	Co$_{3}$\\
\end{tabular}
 & ~\begin{tabular}{c}
	2.00\\
	1.94\\
	1.80\\
\end{tabular}
\\\hline
    S-2 & ~\begin{tabular}{c}
	Pt\\
\end{tabular}
 & ~\begin{tabular}{c}
	0.41\\
\end{tabular}
\\\hline
    S-3 & ~\begin{tabular}{c}
	Co\\
\end{tabular}
 & ~\begin{tabular}{c}
	1.90\\
\end{tabular}
\\\hline
    Center & ~\begin{tabular}{c}
	Pt\\
\end{tabular}
 & ~\begin{tabular}{c}
	0.40\\
\end{tabular}
\\\hline
\end{tabular}
\hspace{1cm}
\begin{tabular}{|c|c|c|c|}
					\hline
	\multicolumn{3}{|c|}{Pt adatom}\\
	\hline
\hline
	Layer & Atom & M ($\mu_{B}$)\\\hline
	Adatom & ~\begin{tabular}{c}
	Pt\\
\end{tabular}
 & ~\begin{tabular}{c}
	0.53\\
\end{tabular}
\\\hline 
	S & ~\begin{tabular}{c}
	Co\\
\end{tabular}
 & ~\begin{tabular}{c}
	2.01\\
\end{tabular}
\\\hline
	S-1 & ~\begin{tabular}{c}
	Pt$_{1}$\\
	Pt$_{2}$\\
	Pt$_{3}$\\
\end{tabular}
 & ~\begin{tabular}{c}
	0.38\\
	0.41\\
	0.35\\
\end{tabular}
\\\hline
    S-2 & ~\begin{tabular}{c}
	Co\\
\end{tabular}
 & ~\begin{tabular}{c}
	1.92\\
\end{tabular}
\\\hline
    S-3 & ~\begin{tabular}{c}
	Pt\\
\end{tabular}
 & ~\begin{tabular}{c}
	0.41\\
\end{tabular}
\\\hline
    Center & ~\begin{tabular}{c}
	Co\\
\end{tabular}
 & ~\begin{tabular}{c}
	1.92\\
\end{tabular}
\\\hline
\end{tabular}
\hspace{1cm}
\begin{tabular}{|c|c|c|c|}
					\hline
	\multicolumn{3}{|c|}{Fe adatom}\\
	\hline
\hline
	Layer & Atom & M ($\mu_{B}$)\\\hline
	Adatom & ~\begin{tabular}{c}
	Fe\\
\end{tabular}
 & ~\begin{tabular}{c}
	3.30\\
\end{tabular}
\\\hline 
	S & ~\begin{tabular}{c}
	Pt$_{1}$\\
	Pt$_{2}$\\
\end{tabular}
 & ~\begin{tabular}{c}
	0.41\\
	0.37\\
\end{tabular}
\\\hline
	S-1 & ~\begin{tabular}{c}
	Co$_{3}$\\
	Co$_{4}$\\
\end{tabular}
 & ~\begin{tabular}{c}
	1.97\\
	1.85\\
\end{tabular}
\\\hline
    S-2 & ~\begin{tabular}{c}
	Pt\\
\end{tabular}
 & ~\begin{tabular}{c}
	0.40\\
\end{tabular}
\\\hline
    S-3 & ~\begin{tabular}{c}
	Co\\
\end{tabular}
 & ~\begin{tabular}{c}
	1.91\\
\end{tabular}
\\\hline
    Center & ~\begin{tabular}{c}
	Pt\\
\end{tabular}
 & ~\begin{tabular}{c}
	0.41\\
\end{tabular}
\\\hline
\end{tabular}\end{center}
\caption{Spin magnetic moments of CoPt thin films on top of which single adatoms are deposited. }
\label{table_ad_atoms}
\end{table*}

After performing the structural relaxations, we investigated the magnetic moments of the films. We found an enhancement of (6.5 \%) and (26.19 \%) of the magnetic moments of Co and Pt adatoms comparatively to those carried by the top-surface layer of the perfect CoPt surface (Co and Pt terminated films respectively). The values of the Co and Fe magnetic moments, respectively 2.13 $\mu_{B}$ and 3.30 $\mu_{B}$, as given in Table \ref{table_ad_atoms} are close to those obtained previously on Pt(111) substrate ~\cite{gambardella2003giant,Balashov2009,Bouhassoune2016}. Interestingly, the moment of Pt adatom increases and reach a value of 0.53 $\mu_{B}$, which is almost the double value obtained for Pt atoms neighbors of adatoms deposited on a Pt substrate.

\begin{figure*}[htpb]
	\centering
	\includegraphics[width=2\columnwidth]{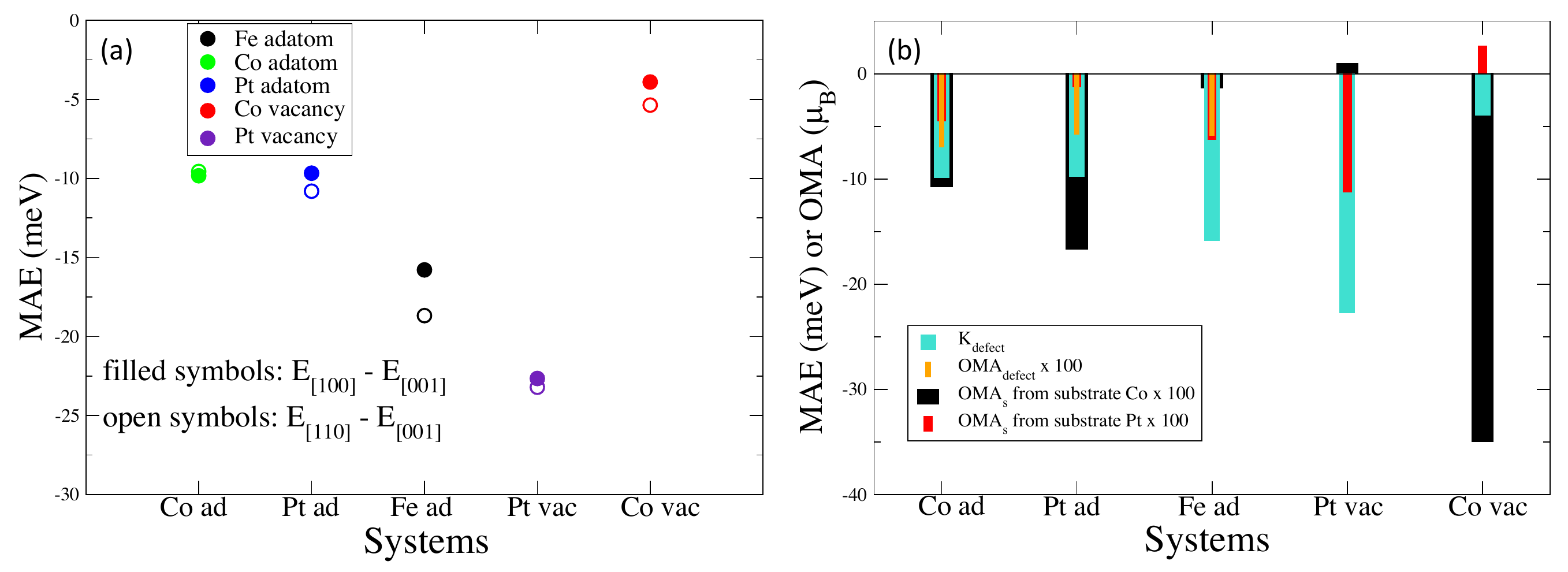} 
	\caption{(a) Contribution of single atoms and vacancies to the total MAE of the investigated unit cells. For completeness, we consider both types of possible in-plane orientations of the moments, along the [110] shown with open symbols and along the [100] direction shown with filled symbols. (b) Impurity-induced MAE and OMA compared to the surface OMA contributions from the Co and Pt layers of the CoPt substrate.}
	\label{mae}
\end{figure*}
\subsection{Magnetic anisotropy energy}
We extract the MAE from total energy differences obtained after rotating the magnetic moments from out-of-plane to in-plane directions, either the [110] or [100] direction. As mentioned earlier, a positive sign of the MAE indicates an out-of-plane preferable orientation of the magnetic moments. We use the following formula to extract the contribution of single atomic defects, being adatoms or surface vacancies, to the total MAE:
\begin{equation}
K_{\mathrm{defect}}=\frac{1}{2}(\mathrm{MAE}^{\mathrm{total}}-\mathrm{MAE}^{\mathrm{defect-free}}),
\end{equation}
where $\mathrm{MAE}^{\mathrm{total}}$ stands for the MAE of the slab including the atomic defect. The factor of $1/2$  originates from the atomic defects being injected on both sides of the slab. We note that $K_{\mathrm{defect}}$ is not necessarily the sole MAE of the defects but it includes the contributions of the substrate reacting to the presence of the impurities.

In Fig.\ref{mae}(a), we plot the contribution of the investigate single atoms and vacancies  to the MAE. As we can see, the difference between the MAEs of the two considered in-plane directions is rather small for Co adatom and Pt vacancy and reaches a maximum for the Fe adatom. For conciseness, we focus the rest of our analysis on a single in-plane direction of the magnetic moments, in particular [100]. Interestingly, we find  that all the defects favor an in-plane orientation of the magnetic moments of the CoPt film. On the one hand, Co adatoms on Pt-terminated surface and Pt adatoms on Co-terminated surface contribute with a practically similar in-plane MAE to the total MAE. Interestingly, one single Pt atom is not enough to switch the MAE to point of plane as it is observed for a full Pt layer covering an initially Co-terminated surface~\cite{brahimi2016giant}. On the other hand, Fe-adatom deposited on Pt-terminated substrate exhibits a much larger in-plane contribution to the MAE.  Among all considered atomic defects, the Pt  (Co) vacancy is the one characterized by the largest (smallest) in-plane contribution to the total MAE. In other words, removing one of the surface Pt atoms basically destroys the out-of-plane orientation of the Pt-terminated CoPt surface by reducing the total MAE by about -22.7 meV (see Fig.\ref{mae}(a)). Our simulations indicate that with Co adatom, the Pt-terminated CoPt thin film maintain its out-of-plane surface MAE, which is not the case for an Fe adatom.  According to these results, this kind of single atomic defects have a dramatic impact on the magnetic orientation of the thin films since the values obtained have to be compared with the defect-free surface contribution to the MAE, which amounts to $\approx$ 16 meV and $\approx$ -4 meV for respectively the Pt-- and Co--terminated surfaces  when the supercells contain 4 atoms per layer~\cite{brahimi2016giant}. 
  Of course, the considered defect concentration is not small in the investigated supercells, 25\%, but these simulations indicate that in the low concentration limit, a similar behavior is expected locally, i.e. we expect regions surrounding the adatoms or vacancies with a tendency to an in-plane orientation of the magnetic moments. This aspect can find practical applications for the manipulation of the MAE.~\cite{Victora2005}

To get further insights on the behavior of the MAE  we use the so-called Bruno's formula ~\cite{Bruno1989}:
\begin{equation}
\mathrm{MAE} = \sum_i \frac{\xi_i}{4}(L^i_{\mathrm{[001]}} - L^i_{\mathrm{[100]}}),
\end{equation}
where $i$ labels the different atoms of the supercells, $\xi$ is the strength of the spin-orbit interaction and $L$ is the orbital magnetic moment calculated when the spin magnetic moment points along 
the [100] or the [001] directions. This model results from perturbation theory assuming that the exchange splitting is larger than the bandwidth. Thus, its validity is probably limited only to the 3$d$ elements like Co and Fe atoms treated in this manuscript. However, it is useful to consider it also for the Pt atoms. The nature of this model permits to relate the MAE to the orbital moment anisotropy (OMA), i.e. $L_{\mathrm{[001]}} - L_{\mathrm{[100]}}$, of each contributing atom and leads to the conclusion that the easy axis of magnetization coincides with the direction having the largest orbital moment. 
Obviously, with this approach we can access  various contributions to the OMA:
\begin{equation}
\mathrm{OMA} = \mathrm{OMA}_{\mathrm{defect}} + \mathrm{OMA}_s + \mathrm{OMA}_b,
\end{equation}
where OMA$_b$ is the bulk contribution and OMA$_s$ corresponds to the surface contribution defined by the change of the OMA because of reduced dimensionality of the slab. We note that in the case of vacancies $\mathrm{OMA}_{\mathrm{defect}} = 0 $.

In Fig.\ref{mae}(b), we plot the different substrate and adatom OMAs contribution to the total OMA compared to $K_\mathrm{defect}$.  Interestingly OMA$_s$ depends drastically on the presence of the defects. We note that except for the case of vacancies, the OMA of the substrate atoms (Co or Pt) as well as of the adatoms favor an in-plane orientation of the moments. Thus the resulting in-plane MAE contribution of the adatoms to the total MAE is a conspiring behavior favored by all elements of the considered CoPt thin films, which experience an in increase of the orbital moments when the spin moments lie in-plane. In the case of the vacancies, the contributions of Co and Pt atoms counter each other. For the Pt vacancy, the Pt contribution to the OMA is rather large and is the largest from all the investigated cases. Considering that the spin-orbit interaction's weight is large for  Pt, this may explain that the Pt-terminated surface with a surface vacancy exhibits the largest in-plane MAE contribution to the total MAE.

\section{Conclusion}
We have investigated from ab-initio the magnetic properties of adatoms and vacancies on the surface of tetragonal {\bf{L1}$_{0}$} CoPt thin films. After structural relaxations determining the preferable sites of Co, Fe and Pt adatoms, we explored the magnetic properties of the substrate and defects. While Co and Pt adatoms prefer the hollow site, Fe adatom  favors the bridge site. The surface and subsurface atoms, to a lower extent, rearrange themselves to accommodate the defects, which affects the spin and orbital magnetic moments. This impacts the anisotropy of the orbital magnetic moments of the defects and substrate, which is maximized when the spin magnetic moments point in-plane. Within perturbation theory this explains that the defects contribute by an in-plane MAE to the total MAE. This result is in line with our previous findings that the out-of-plane MAE of {\bf{L1}$_{0}$} CoPt  is sensitive to defects~\cite{brahimi2016giant}.

\section*{Acknowledgments}
We are grateful to Vasile Caciuc, Rico Friedrich, Julen Ibanez-Azpiroz and Manuel dos Santos Dias for helpful discussions.  The simulations were performed on the cluster of computers "Aselkam" of the University Mouloud Mammeri in Tizi-Ouzou. 
This work was supported by C. N. E. P. R. U project (D 00520090041) of the Algerian government and the European Research
Council (ERC) under the European Union's Horizon 2020 research and
innovation programme (ERC-consolidator Grant No. 681405 DYNASORE).

\end{document}